\documentclass[12pt]{iopart}
\pdfoutput=1
\usepackage{citesort}
\usepackage{iopams}
\usepackage{graphicx}
\usepackage{setstack}
\begin{document}

\title[]{Edge states of hydrogen terminated monolayer materials: silicene, germanene and stanene ribbons}
\author{A Hattori$^1$, S Tanaya$^2$, K Yada$^1$, M Araidai$^{3,4}$, M Sato$^{5}$, Y Hatsugai$^2$, K Shiraishi$^3$, Y Tanaka$^1$}
\address{$^1$ Department of Applied Physics, Nagoya University, Nagoya 464-8603, Japan}
\address{$^2$ Division of Physics, Faculty of Pure and Applied Sciences, University of Tsukuba, Tsukuba, Ibaraki 305-8571, Japan}
\address{$^3$ Institute of Materials and Systems for Sustainability, 
Nagoya University, Nagoya 464-8603, Japan}
\address{$^4$ Institute for Advanced Research, Nagoya University, Nagoya 464-8601, Japan}
\address{$^5$ Yukawa Institute for Theoretical Physics, Kyoto University, Kyoto 606-8502, Japan}
\ead{hattori@rover.nuap.nagoya-u.ac.jp} 

\begin{abstract}

We investigate the energy dispersion of the edge states in zigzag silicene, germanene and 
stanene nanoribbons with and without hydrogen 
termination based on a multi-orbital tight-binding model.  Since the low buckled 
structures are crucial for these materials, both the $\pi$ and $\sigma$ 
orbitals have a strong influence on the edge states, different from the case for graphene 
nanoribbons. The obtained dispersion of helical edge states is nonlinear, similar to that obtained by 
first-principles calculations. On the other hand, the dispersion derived from the 
single-orbital tight-binding model is always linear. Therefore, we find that the 
non-linearity comes from the multi-orbital effects, and accurate results cannot be obtained 
by the single-orbital model but can be obtained by the multi-orbital tight-binding model. We show that 
the multi-orbital model is essential for correctly understanding the dispersion of the edge 
states in tetragen nanoribbons with a low buckled geometry. 

\end{abstract}

\vspace{2pc}
 \par
\submitto{\JPCM} 
\noindent (Some figures may appear in colour only in the online journal)

\maketitle
%\ioptwocol

\section{Introduction}
Silicene, germanene and stanene are quasi-two dimensional 
graphene-like materials composed of  
silicon, germanium and tin atoms, respectively. 
Graphene has massless Dirac cones at the $K$ and $K'$ points in the Brillouin zone at which the $\pi$ and $\pi^{\ast}$ bands 
linearly cross the Fermi level. 
The electronic properties of silicene, germanene and stanene are akin to 
those of graphene;
however, the Dirac cones become  
massive due to spin--orbit coupling \cite{takedashiraishi,2dsige,4bulk,Ezawa_review}. 
These three materials prefer to construct $sp^{3}$-like hybridized 
orbitals rather than $sp^{2}$ ones and exhibit low buckled structures, different  
from that of graphene. 
Owing to this low buckled structure, these three materials have
ambipolar properties and the magnitude of the energy gap is tunable by 
an electric field \cite{silgerQSH,electric,Ezawa_externalfield}. 
For these materials, 
quantum spin Hall (QSH) effects 
are predicted and can be observed at room temperature in stanene  \cite{Ezawa_VPM,stanene}. 
Silicene sheets have been synthesized on 
Ag \cite{Ag1,Ag2,Ag3,Ag4,Ag5,sildevice}, ZrB$_2$ \cite{Zr} and Ir \cite{Ir} substrates under ultra-high pressure vacuum \cite{support}, 
and it has been reported that 
germanene has been synthesized on substrates 
of Au and Pt \cite{Au,Pt} and  
stanene has been synthesized on Bi$_{2}$Te$_{3}$ \cite{Bi2Te3}. \par

Several studies of tetragen nanoribbons have been reported. 
For graphene nanoribbons (GrNRs), 
a flat band zero energy edge state is generated from 
the $K$ point to the zone boundary in a zigzag edge and 
from the $\Gamma$ to $K$ points in a Klein edge  \cite{gra1,graokaoshi,gra2,gra3,hatsu1,hatsu2}. 
In actual experiments, the hydrogen termination effect 
is important as it  
has a strong influence on the edge states of GrNRs  \cite{edge_ex,hydrogra4,hydrogra3}. 
Zigzag graphene nanoribbons (ZGrNRs) with mono-hydrogen termination 
at the edge sites reproduce the zigzag edge states \cite{graribbonex1,edge_oka,graribbonex4,edge_yama}. 
On the other hand, ZGrNRs with di-hydrogen termination at the  edge sites 
reproduce the Klein edge state \cite{hydrogra1,hydrogra2}.

The electronic properties of the edge states of silicene, germanene 
and stanene nanoribbons (SiNRs, GeNRs, SnNRs) 
with and without hydrogen termination 
have also been studied using a single-orbital tight-binding model  \cite{Ezawa_externalfield,Ezawa_review}
and first-principles calculations   \cite{hydro_passivated_sil,germanene_nanorod,functionalized_germanene,decorate_sn,snonge} 
by many groups. 
Although the single-orbital model can reproduce the bulk energy dispersion  \cite{effective}, 
it has not yet been clarified whether the single-orbital model can really express 
the various types of edge states with hydrogen termination \cite{Ezawa_review}. 
On the other hand, 
it has been reported based on first-principles calculations 
that the single-orbital model cannot correctly express the dispersion of hydrogen 
terminated edge states 
of zigzag germanene nanoribbons (ZGeNRs) \cite{fujitager}. 
Although the single-orbital tight-binding model is too simple to reproduce the correct edge state, 
it is not clear whether a multi-orbital model can correctly express the hydrogen terminated edge states. 

In this paper, we construct multi-orbital tight-binding models for 
zigzag silicene nanoribbons (ZSiNRs), ZGeNRs and zigzag stanene nanoribbons (ZSnNRs). 
The parameters are extracted from the density-functional theory (DFT)  results for simple tetragen molecules. We 
show that the energy dispersion derived from the DFT calculations can be successfully reproduced 
by our multi-orbital tight-binding model with the parameters. 
In the case of mono-hydrogen termination, zigzag-like edge states are realized, 
while in the di-hydrogen termination case, Klein-like edge states are realized. 
We obtain the nonlinear dispersion of the edge states, 
which is different from that based on a single-orbital model 
and is consistent with the results of first-principles calculations \cite{fujitager}.
Based on these results, we find that the origin of this nonlinear dispersion is 
the $sp^{3}$-like hybridized orbitals. 
This shows that the multi-orbital tight-binding model is useful for studying the electronic properties 
of nanoribbons including the edge states.  

The organization of this paper is as follows. 
In \Sref{model}, we explain the model and formulation 
as well as the structure of the nanoribbons, the multi-orbital tight-binding model and 
the material parameters. 
We also explain 
how the coupling constants for Si-H, Ge-H, Sn-H 
bonds and the on-site energies of each orbital are determined.
In \Sref{Result}, we show the results of our numerical calculations, 
including the energy spectra of zigzag nanoribbons (ZNRs) with and without hydrogen termination. 
In \Sref{conclusion}, we summarize our results.

\section{Models and Formulations\label{model}}

\subsection{Atomic structure  of nanoribbons}
\begin{figure*}[htbp]
	\includegraphics[width=150mm]{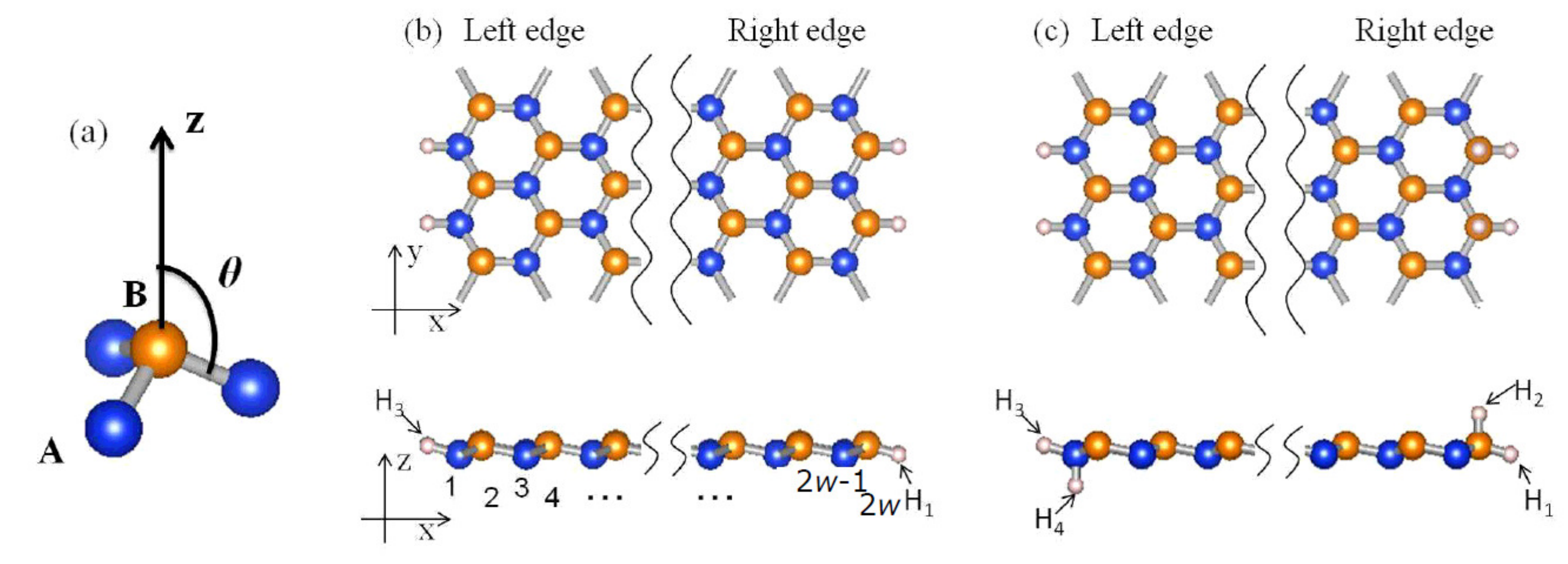}
	\caption{\label{fig:model}(a) Illustration of $\theta$, defined as 
being the angle between the direction from the B site (orange) to the A site (blue) and the 
$z$-axis normal to the plane. (b), (c) Lattice geometry of ZNRs with hydrogen termination at both edge sites: (b) 1H/1H and (c) 2H/2H. The upper and lower panels show the top and side views.}
\end{figure*}

SiNRs, GeNRs and SnNRs have low buckled honeycomb structures, as shown in \Fref{fig:model}.
The buckling angle $\theta$ is defined as shown in \Fref{fig:model}(a).
The vectors from the B site to the three neighboring A sites ($\vec{d_1}$, $\vec{d_2}$ and $\vec{d_3}$) are given by
	\begin{equation}
		\vec{d_1}=a\sin{\theta}(1,0,\cot{\theta}), 
	\end{equation}
	\begin{equation}
		\vec{d_2}=a\sin{\theta}(-\frac{1}{2},\frac{\sqrt{3}}{2},\cot{\theta}), 
	\end{equation}
	\begin{equation}
		\vec{d_3}=a\sin{\theta}(-\frac{1}{2},-\frac{\sqrt{3}}{2},\cot{\theta}),
	\end{equation}
with a lattice constant $a$. 
Here we consider nanoribbon systems 
which have periodic boundaries along the $y$-axis, as shown in Figures \ref{fig:model}(b) and (c). 
The width of the nanoribbons is denoted by $2w$. 
Figures \ref{fig:model}(b) and (c) show two kinds of hydrogen termination (mono-hydrogen and di-hydrogen termination). 
Considering the $sp^3$ nature of tetragens, we choose the position vectors of the hydrogen site from the tetragen site at the rightmost edge to be
	\begin{equation}
		\vec{d_{\mathrm{H}_1}}=a_\mathrm{H}(\sin{\theta},0,\cos{\theta}),
	\end{equation}
	\begin{equation}
		\vec{d_{\mathrm{H}_2}}=a_\mathrm{H}(0,0,1),
	\end{equation}
where $a_\mathrm{H}$ is the distance between a hydrogen and a tetragen.
The azimuthal angle $\theta$ is chosen to be the same as the buckling angle.
The position vectors for the left edge are opposite to those of the rightmost edge, i.e., $\vec{d_{\mathrm{H}_3}}=-\vec{d_{\mathrm{H}_1}}$ and $\vec{d_{\mathrm{H}_4}}=-\vec{d_{\mathrm{H}_2}}$. 
In this paper, we consider three kinds of hydrogen termination:
mono-hydrogen termination at the outermost edges (1H/1H), 
di-hydrogen termination at the outermost edges (2H/2H) 
and raw edges without hydrogen termination (0H/0H). 

\subsection{Multi-orbital tight-binding model}
The multi-orbital tight-binding model used in this paper consists of four outer-shell
orbitals ($s$, $p_x$, $p_y$ and $p_z$ orbitals) of tetragens and a $1s$-orbital of hydrogen.
The Hamiltonian $\mathcal{H}$ is given by 
	\begin{equation}
		\mathcal{H}=\mathcal{H}_0+\mathcal{H}_{\mathrm{so}}+\mathcal{H}_{\mathrm{H}}
		.\label{eq1} 
	\end{equation}
The first term $\mathcal{H}_0$ denotes the on-site energy of the tetragens and nearest neighbor hopping between them: 
	\begin{eqnarray}
			\mathcal{H}_0=&\sum_{\langle i,j \rangle} \sum_{\alpha,\beta} \sum_{\tau} (t^{\alpha\beta}_{i,j} c^\dagger_{i\alpha \tau}c_{j\beta \tau}+\mathrm{h.c.}) \nonumber \\
			&+\sum_{i} \sum_{\alpha} \sum_{\tau} \epsilon_\alpha c^\dagger_{i\alpha\tau}c_{i\alpha\tau},\label{eq7}
	\end{eqnarray}
where $c^\dagger_{i\alpha \tau}$ and $c_{i\alpha \tau}$ are the creation and annihilation
operators for an electron with an atomic orbital $\alpha$ and a spin $\tau$ at site $i$. $\epsilon_{\alpha}$ denotes the site energy for orbital $\alpha$. 
The first term in equation \eref{eq7} corresponds to the hybridization  
between tetragens and the second term represents the on-site energy at the tetragen sites. 
$\langle i,j \rangle$ runs over all the nearest neighbor hopping sites. 
The hopping integral $t^{\alpha\beta}_{i,j}$ is determined by the Slater--Koster parameters $V_{ss\sigma}$, $V_{sp\sigma}$, $V_{pp\sigma}$ and $V_{pp\pi}$,
as shown in \Tref {tab:slater}. 
The second term $\mathcal{H}_{\mathrm{so}}$ expresses the spin--orbit interaction 
	\begin{equation}
		\mathcal{H}_{\mathrm{so}}=\frac{\xi_0}{2}\sum_{i} \sum_{\bar{\alpha}\bar{\beta}\bar{\gamma}} \sum_{\tau,\tau^{\prime}}  \epsilon_{\bar{\alpha}\bar{\beta}\bar{\gamma}}c^{\dagger}_{i\bar{\alpha} \tau}(-i\hat{\sigma}_{\bar{\gamma}}) c_{i\bar{\beta} \tau^{\prime}}+\mathrm{h.c.}, 
	\end{equation} 
where $\xi_0$ is the strength of the spin--orbit coupling, 
$\bar{\alpha}=x,y,z $, $\bar{\beta}=x,y,z$ and $\bar{\gamma}=x,y,z$ are indices of  the $p_{\bar{\alpha}}$, $p_{\bar{\beta}}$ and $p_{\bar{\gamma}}$ orbitals.  
$\epsilon_{\bar{\alpha}\bar{\beta}\bar{\gamma}}$ is an antisymmetric tensor and  
$\hat{\sigma}_{\bar{\gamma}}$ is the Pauli matrix  acting on the spin space.
The third term of $\mathcal{H}_\mathrm{H}$ describes the hydrogen termination: 
	\begin{eqnarray}
		\mathcal{H}_\mathrm{H}=&\sum_{\langle i,j \rangle} \sum_\alpha \sum_{\tau} (t^{s\alpha}_{ij} d^{\dagger}_{is\tau} c_{j\alpha\tau}+\mathrm{h.c.}) \nonumber \\
		&+\sum_i \sum_{\tau} \epsilon_H d^{\dagger}_{is\tau} d_{is\tau},\label{eq9}
	\end{eqnarray}
where $d^\dagger_{is\tau}$ and $d_{is\tau}$) are the creation and annihilation operators for an electron at hydrogen site $i$. 
The first term in equation \eref{eq9} corresponds to the hybridization of 
hydrogen and tetragen, and the second term is the on-site energy at the hydrogen sites. 
$\epsilon_{H}$ denotes the site energy for an electron at a hydrogen atom. 
The hopping parameters for the tetragens' bonds, i.e., the values of $V_{ss\sigma}$, $V_{sp\sigma}$, $V_{pp\sigma}$ and $V_{pp\pi}$, 
are given in \Tref{tab:value1}.
The hopping parameters between tetragens and hydrogens are given in \Tref{tab:value2}, 
which will be explained in \Sref{hydrogen}. 

\begin{table}
\caption{\label{tab:slater}Slater--Koster interatomic matrix 
elements \cite{slater}. The matrix elements for nearest neighbor 
hopping between $s$ and $p$ orbitals are determined by the 
direction cosines $l_{ij}$, $m_{i,j}$ and $n_{i,j}$ which are the $x$, 
$y$ and $z$ components measured from site $i$ to site $j$.}  

\begin{indented}
\lineup
\item[]\begin{tabular}{@{}*{2}{c}}
\br                              
$t^{ss}_{ij}$     &$V_{ss\sigma}$\cr 
\mr
$t^{sx}_{ij}$     &$l_{ij}V_{sp\sigma}$ \cr
\mr
$t^{xs}_{ij}$    &$-l_{ij}V_{sp\sigma}$\cr 
\mr
$t^{xx}_{ij}$   &  $l^2_{ij}V_{pp\sigma}+(1-l^2_{ij})V_{pp\pi}$\cr 
\mr
$t^{xy}_{ij}$   &  $l_{ij}m_{ij}(V_{pp\sigma}-V_{pp\pi})$\cr
\mr
$t^{xz}_{ij}$    & $m_{ij}n_{ij}(V_{pp\sigma}-V_{pp\pi})$\cr 
\br
\end{tabular}
\end{indented}
\end{table}

\begin{table*}
\caption{\label{tab:value1} Numerical values of parameters in multi-orbital tight-binding models. 
The lattice constants $a$ and the buckling angles $\theta$ are given by Liu \etal.  \cite{effective}. 
The hopping parameters between tetragens are chosen according to 
Ref.~\cite{c_bond} for C-C bonds, Ref.~\cite{sige_bond} for Si-Si and Ge-Ge bonds 
and Ref.~\cite{sn_bond} for Sn-Sn bonds. The strength of the spin--orbit coupling 
$\xi_{0}$ for graphene, silicene (germanene) and stanene are obtained based on Yao 
\etal. \cite{c_soc}, Liu \etal. \cite{silgerQSH} and Chadi \cite{sn_soc}.} 
		
\begin{indented}
\lineup
\item[]\begin{tabular}{@{}*{8}{c}}
\br
System              &  $a$(\AA)   &  $\theta$(deg)  &  $V_{ss\sigma}$(eV)  &  $V_{sp\sigma}$(eV)  &  $V_{pp\sigma}$(eV)  &  $V_{pp\pi}$(eV)             & $\xi_0$(eV)  \cr 
\mr
Graphene     & 2.46         & 90           & $-6.769$                                & 5.580             & 5.037                     & $-3.033$                             & $9\times10^{-3}$           \cr
\mr
Silicene     & 3.86         & 101.7           & $-1.93$                               & 2.54             & 4.47                     & $-1.12$                             & $34\times10^{-3}$           \cr
\mr
Germanene & 4.02         & 106.5          & $-1.79$                                & 2.36                               & 4.15                   & $-1.04$             &  0.196           \cr
\mr
Stanene     & 4.70        & 107.1           & $-2.6245$                              & 2.6504                           & 1.4926                & $-0.7877$         &  0.8                 \cr
\br
\end{tabular}
\end{indented}
\end{table*}

\begin{table*}
\caption{\label{tab:value2} Parameters of C-H, Si-H, Ge-H and Sn-H bonds 
and the difference in on-site energies (eV) in the multi-orbital tight-binding model.} 

\begin{indented}
\lineup
\item[]\begin{tabular}{@{}*6{c}} 
\br 
System      &  $V_{ss\sigma}$  &  $V_{sp\sigma}$  &  $\epsilon_{s}$ & $\epsilon_s-\epsilon_H$  &  $\epsilon_s-\epsilon_p$  \cr
\mr
Graphene     & $-10.457$ \cite{tanaya}& 13.744 \cite{tanaya}&  $-17.52$ \cite{tanaya}& $-3.87$ \cite{Harrison,tanaya}&  $-8.55$ \cite{Harrison} \cr 
\mr
Silicene     & $-3.18$        & 3.32           & $-7.90$& $-1.97$         &  $-5.44$   \cr
\mr
Germanene & $-3.29$         & 2.66          &$-7.90$& $-1.00$          & $-6.74$   \cr
\mr
Stanene     & $-2.75$        & 3.27           &$-9.00$& $-4.38$          & $-5.61$    \cr
\br
\end{tabular}
\end{indented}
\end{table*}

\subsection{Low-energy effective Hamiltonian}
In this subsection, we explain the single-orbital model which effectively 
expresses the low-energy dispersion around the $K$ and $K'$ points \cite{effective}. 
The Hamiltonian  $\mathcal{H}_{\mathrm{eff}}$ in the single-orbital model is given by 
	\begin{eqnarray}
				\mathcal{H}_{\mathrm{eff}}=&-t\sum_{\langle i,j \rangle} \sum_{\tau} c^\dagger_{i\tau}c_{j\tau}+i\frac{\lambda_{\mathrm{so}}}{3\sqrt{3}}\sum_{\langle \langle i,j \rangle \rangle}  \sum_{\tau\tau^{\prime}} v_{ij}c^\dagger_{i\tau}\sigma^z_{\tau\tau^{\prime}}c_{j\tau'} \nonumber \\
				&-i\frac{2}{3}\lambda_\mathrm{R} \sum_{\langle \langle i,j \rangle \rangle} \sum_{\tau\tau^{\prime}} \mu_i c^\dagger_{i\tau}{(\sigma\times\hat{d}_{ij})}^{z}_{\tau\tau^{\prime}} c_{j\tau^{\prime}}, 
	\end{eqnarray}
where $c^\dagger_{i\tau}$ and $c_{i\tau}$ are creation and annihilation
operators with spin $\tau$ at site $i$. 
The first term represents nearest neighbor hopping. 
The second and third terms are the effective spin--orbit coupling 
and the intrinsic Rashba spin--orbit coupling, respectively. 
$\langle \langle i,j \rangle \rangle$ runs over all the next nearest neighbor sites. 
As shown by Kane and Mele \cite{kanemele}, 
$v_{i,j}=+1$ or $-1$ if the direction from $j$ to $i$ site is anticlockwise or clockwise in the hexagon. 
$\sigma^{z}_{\tau\tau^{\prime}}$ denotes an element of 
the Pauli matrix acting on the spin space.
$\mu_i=1$ and $-1$ for the A or B site and $\hat{d}_{i,j}$ is given by 
$\hat{d}_{i,j}=\vec{d}_{i,j}/|\vec{d}_{i,j}|$, where $\vec{d_{i,j}}$ is a vector from site $i$ to the second nearest neighbor site $j$. 
The material parameters for a single-orbital Hamiltonian are shown in \Tref{tab:effective} \cite{Ezawa_review,effective}. 
Here, $t$ and $\lambda_{\mathrm{so}}$ are based on a paper by Liu \etal.  \cite{effective}.
\begin{table}
\caption{\label{tab:effective}Material parameters in effective model  \cite{effective,Ezawa_review}. }
	
\begin{indented}
\lineup 
\item[]\begin{tabular}{ccccc} 
\br
System      &  $t$ (eV)  &  $\lambda_{so}$ (meV) & $\lambda_R$ (meV)  &  $l$ (\AA)  \cr
\mr
Graphene   &  2.8      &   10$^{-3}$   &  0  &  0  \cr
\mr
Silicene     &  1.07         & 3.97        &  0.7   &  0.23   \cr
\mr
Germanene & 0.991         & 46.3          & 10.7  &  0.33  \cr
\mr 
Stanene     & 0.760        & 64.4          & 9.5   & 0.40   \cr
\br
\end{tabular}
\end{indented}
\end{table}

\subsection{
Determination of parameters for Si-H, Ge-H and Sn-H bonds 
and on-site energies\label{hydrogen}}

\begin{figure*}[tbp]
	\includegraphics[width=12cm]{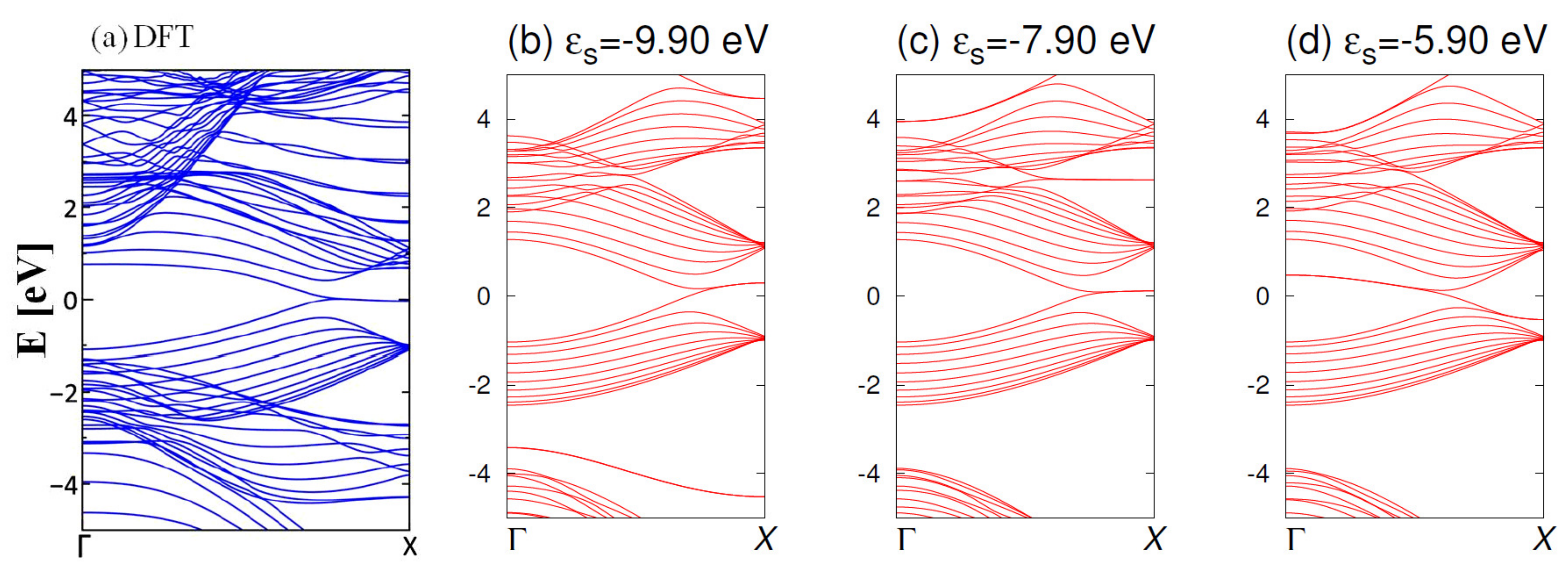}
	\caption{\label{fig:1h1hsilon} Energy spectra for 1H/1H ZSiNRs ($w=10$). (a) First-principles calculation. \cite{first} (b), (c), (d) Calculation using multi-orbital tight-binding model: 
(b) $\epsilon_{s}=-9.90$ eV, (c) $\epsilon_{s}=-7.90$ eV, 
(d)$\epsilon_{s}= -5.90$ eV.}
\end{figure*}

In this subsection, 
we explain how to determine the values of the binding energies for Si-H, Ge-H and Sn-H bonds and the on-site energy of each orbital in \Tref{tab:value2}. 
For this purpose, we need the interatomic elements for coupled silicon, germanium and tin and hydrogen atoms.
We determine these elements 
from molecules that have Si-H, Ge-H and Sn-H bonds. 
In this paper, we use the simple molecules silane (SiH$_4$), germane (GeH$_4$) and stannane (SnH$_4$).
For these molecules, we assume a multi-orbital Hamiltonian $\mathcal{H}_{\mathrm{mol}}$ with nearest neighbor hopping. 
In the case of silane, the relevant orbitals 
are \{3$s$, 3$p_x$, 3$p_y$, 3$p_z$, 1$s$, 1$s$, 1$s$, 1$s$\}. 
$\mathcal{H}_{\mathrm{mol}}$ is given by 
\begin{eqnarray}
		\mathcal{H}_{\mathrm{mol}}=&\sum_{\langle i,j \rangle} \sum_{k=0}^{3} \sum_\alpha \sum_{\tau} (t^{s\alpha}_{ij} {d^{(k)}}^{\dagger}_{is\tau} c_{j\alpha\tau}+\mathrm{h.c.}) \nonumber \\
		&+\sum_i \sum_{k=0}^3	\sum_{\alpha} \sum_{\tau} (\epsilon_H {d^{(k)}}^{\dagger}_{is\tau} d^{(k)}_{is\tau}+\epsilon_\alpha c^{\dagger}_{i\alpha\tau} c_{i\alpha\tau}). \nonumber \\
{}		
\end{eqnarray}
From this 8$\times$8 Hamiltonian, 
we can analytically obtain eight eigenvalues  which depend on the interatomic elements ($V_{ss\sigma}$, $V_{sp\sigma}$) 
and on-site energies ($\epsilon_s$, $\epsilon_p$, $\epsilon_H$),  
 as shown in Appendix A.  
The four obtained energy eigenvalues have a single 
degeneracy and triple degeneracies 
at the HOMO and LUMO levels, respectively. 
They are given by 
\begin{equation}
	\lambda^\pm_1=\frac{\epsilon_H+\epsilon_s\pm\sqrt{(\epsilon_H-\epsilon_s)^2+16V^2_{ss\sigma}}}{2},\label{eq13}
\end{equation}
\begin{equation}
	\lambda^\pm_3=\frac{\epsilon_H+\epsilon_p\pm\sqrt{(\epsilon_H-\epsilon_p)^2+16V^2}}{2},\label{eq14}
\end{equation}
with $V=V_{sp\sigma}/\sqrt{3}$,
 where $\lambda^\pm_1$ has a single degeneracy and 
$\lambda^\pm_3$ have triple degeneracies. 
The values of $\lambda^\pm_1$ and $\lambda^\pm_3$ 
are determined by first-principles calculations, as shown in \Tref{tab:level} \cite{first}. 
Then, by using 
equations \eref{eq13} and \eref{eq14}, we determine five parameters while the number of equations is four.  
Thus, we first obtain a value of $\epsilon_{s}$ as a parameter. 
Then the other four parameters are given by the following expressions: 
\begin{equation}
					\epsilon_p=\epsilon_s-(\lambda^+_1+\lambda^-_1)+(\lambda^+_3+\lambda^-_3), 
				\end{equation}
				\begin{equation}
					\epsilon_H=-\epsilon_s+(\lambda^+_1+\lambda^-_1),
				\end{equation}
				\begin{equation}
					V_{ss\sigma}=\frac{\sqrt{(\lambda^+_1-\lambda^-_1)^2-(\epsilon_H-\epsilon_s)^2}}{4},\label{eq17}
				\end{equation}
				\begin{equation}
					V=\frac{\sqrt{(\lambda^+_3-\lambda^-_3)^2-(\epsilon_H-\epsilon_p)^2}}{4}.\label{eq18}
				\end{equation}
equations  \eref{eq17} and \eref{eq18} require 
$(\lambda^{+}_{1}-\lambda^{-}_{1})^{2}-(\epsilon_{H}-\epsilon_{s})^{2} > 0$ 
and $(\lambda^+_3-\lambda^-_3)^2-(\epsilon_H-\epsilon_p)^2 > 0$, 
giving the following constraint 
	\begin{eqnarray}
	\left\{
	\begin{array}{c}
		\lambda^-_1<\epsilon_s<\lambda^+_1 \\ 
		\lambda^+_1+\lambda^-_1-\lambda^+_3<\epsilon_s<\lambda^+_1+\lambda^-_1-\lambda^-_3. 
	\end{array}
	\right. \label{eq19}
	\end{eqnarray}
Using equation \eref{eq19}, we can determine the values of  $\epsilon_s$ to be
for Si: $-13.4<\epsilon_s<-5.43$, Ge: $-14.0<\epsilon_s<-6.57$ 
and Sn : $-12.7<\epsilon_s<-5.79$. 
Substituting these material parameters into the Hamiltonian for the ZNRs $\mathcal{H}$ 
given in  equation \eref{eq1},
we can calculate the energy spectra for ZNRs with hydrogen termination at 
the outermost edge sites 
and compare with those obtained by first-principles calculations for $w=10$ \cite{first}. 
We compare the energy spectra obtained by the multi-orbital model with 1H/1H and 2H/2H 
ZNRs and those by first-principles calculations, and optimize the value of 
$\epsilon_s$ for ZSiNRs, ZGeNRs and ZSnNRs. 
\Fref{fig:1h1hsilon} shows energy spectra for 1H/1H ZSiNRs using 
(a) first-principles calculations  
and (b)--(d) the multi-orbital tight-binding model. 
We have chosen $\epsilon_s=-7.90$ eV  since 
we can qualitatively reproduce 
the energy spectra obtained by first-principles calculations. 
 We can also reproduce the energy spectra for 2H/2H ZSiNRs obtained by first-principles calculations. 
Similarly, we have chosen values of $\epsilon_{s}$ for ZGeNRs and ZSnNRs as summarized in \Tref{tab:value2}. 

\begin{table}
\caption{\label{tab:level}Energy levels (eV) of SiH$_4$, GeH$_4$ and SnH$_4$ obtained by first-principles calculations \cite{first}. } 

\begin{indented}
\lineup
\item[]\begin{tabular}{ccccc} \cr
\br
System      &  $\lambda^+_3$  &  $\lambda^+_1$  &  $\lambda^-_3$  &  $\lambda^-_1$  \cr
\mr
Silane     & 0.00880         &   $-0.475$     & $-8.40$                         &  $-13.4$    \cr
\mr
Germane & 0.169         & $-0.790$          & $-8.23$                      & $-14.0$     \cr
\mr
Stannane     & $-0.181$        & $-0.882$           & $-7.82$                    & $-12.7$   \cr
\br
\end{tabular}
\end{indented}
\end{table}

\begin{table}
\caption{\label{tab:sih}Numerically obtained parameters for 
Si-H (eV).}

\begin{indented}
\lineup
\item[]\begin{tabular}{cccc} \cr
\br
$\epsilon_s$                      &   $-9.90$         &   $-7.90$        &    $-5.90$ \cr
\mr
$\epsilon_s$-$\epsilon_p$   &   $-5.44$   &   $-5.44$  &  $-5.44$  \cr
\mr
$\epsilon_s$-$\epsilon_H$   &   $-5.97$   &   1.62   &   2.03  \cr
\mr
$V_{ss\sigma}$                  &$-2.85$    &   $-3.18$     &  $-3.18$  \cr
\mr
$V_{sp\sigma}$                 &   3.64   &   3.32       &   1.67   \cr
\br
\end{tabular}
\end{indented}
\end{table}

\section{Results \label{Result}}
\subsection{ZNRs without hydrogen termination}
In this subsection, 
we calculate the energy spectra for ZNRs without hydrogen termination for $w=100$.
By comparing the results obtained by the single-orbital model with those 
by the multi-orbital model, we demonstrate the benefit of 
using the multi-orbital model to study the energy spectra for ZNRs 
having low buckled geometries. 

\begin{figure}[htbp]
	\includegraphics[width=7cm]{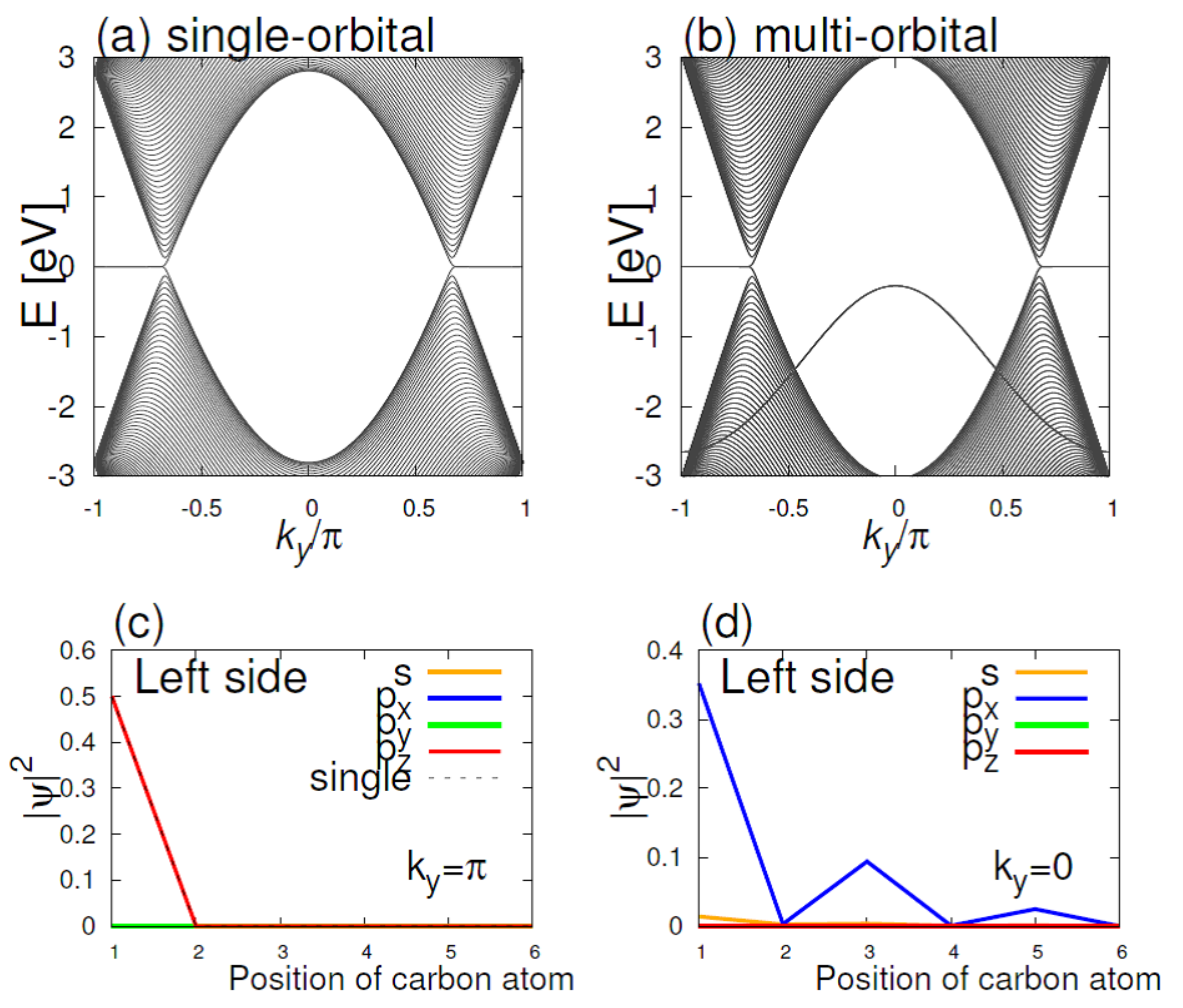}
	\caption{\label{fig:0h0hgra} Energy spectra for ZGrNRs ($w=100$) calculated by (a) the single-orbital model and (b) the multi-orbital model (0H/0H). Orbital decomposed probability density of edge states 
(c) $k_{y}=\pi$ and (d) $k_{y}=0$ in 0H/0H ZGrNRs.}
\end{figure}

First, we show the energy spectra for ZGrNRs based on the (a) single-orbital and (b) multi-orbital models 
in \Fref{fig:0h0hgra}. 
In the figure, 
both the single-orbital model and the multi-orbital model exhibit 
a flat band  in the momentum space  $2\pi/3 \leq |k_{y}| \leq \pi$ at the Fermi level. 
The energy spectrum for the single-orbital model qualitatively 
reproduces that for the multi-orbital model 
except for the in-gap states at $-0.4\pi \leq k_y \leq 0.4 \pi$,
shown in \Fref{fig:0h0hgra}(b). 
Figures \ref{fig:0h0hgra}(c) and (d) show the orbital decomposed probability density $|\Psi(\alpha)|^{2}$ ($\alpha=s, p_{x}, p_{y}, p_{z}$) of the flat band and in-gap states, respectively. 
$|\Psi(p_{z})|^{2}$ of the flat band at $k_{y}=\pi$ 
is localized only at the outermost edge sites. 
This feature agrees with that for the single-orbital model as shown in \Fref{fig:0h0hgra}(c).  
Therefore, the flat band in Figures \ref{fig:0h0hgra}(a) and (b) corresponds to the zigzag edge states.
On the other hand, the in-gap states at   
$-0.4\pi \leq k_{y} \leq 0.4 \pi$ in the multi-orbital model 
mainly consist of the $p_{x}$ orbital around the edge sites of ZGrNRs, 
and $|\Psi(p_{x})|^{2}$ oscillates with 
damping from the outermost edge sites to the inner sites 
as shown in \Fref{fig:0h0hgra}(d). 
Since the $\sigma$ orbitals composed of 2$s$, 2$p_{x}$ and 2$p_{y}$ components are orthogonal to the 2$p_{z}$ ($\pi$) orbital in ZGrNRs, 
the flat band originates only from the $\pi$-bonds. 
Then, without considering the multi-orbital effect, non-zero edge states 
do not appear. 
The non-zero edge state is a dangling bond on a carbon atom at the edge site of ZGrNRs. 
The carbon 2$s$, 2$p_{x}$ and 2$p_{y}$ orbitals are essential for describing both the dangling bonds and the orbital  hybridizations between carbon and hydrogen,  
as shown in the next subsection. 

\begin{figure}[htbp]
	\includegraphics[width=7cm]{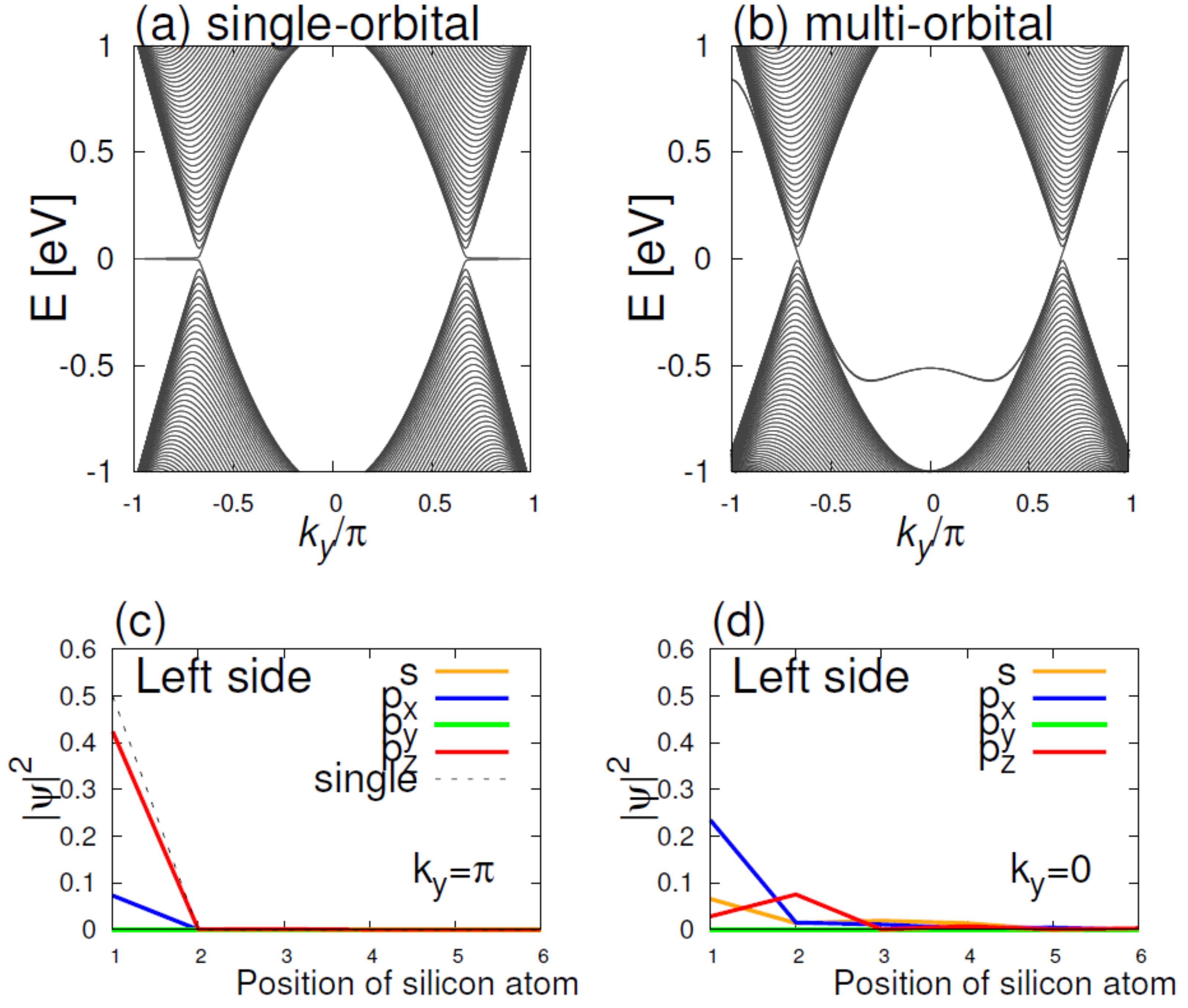}
	\caption{\label{fig:0h0hsil} Energy spectra for ZSiNRs ($w=100$) calculated by (a) the single-orbital model and (b) the multi-orbital model (0H/0H). Orbital decomposed probability density 
of edge states  at (c) $k_{y}=\pi$, (d) $k_{y}=0$ in 0H/0H ZSiNRs.}
\end{figure}

Next, we show 
the energy spectrum for ZSiNRs without hydrogen termination.
As seen from \Fref{fig:0h0hsil},
the bulk energy spectra  for the single-orbital model (\Fref{fig:0h0hsil}(a)) and the multi-orbital model (\Fref{fig:0h0hsil}(b))
are similar.  
However while there exists a nearly flat band in the single-orbital model similar to the graphene case (Figures \ref{fig:0h0hgra}(a) and (b)),   
there is no such band in the multi-orbital model, as shown in \Fref{fig:0h0hsil}(b). 
In the multi-orbital model, instead of a flat band, in-gap states with dispersion  
appear for $2\pi/3\leq|k_{y}|\leq\pi$ with positive energy and 
for  $-2\pi/3 \leq k_{y} \leq 2\pi/3$ with negative energy. 
We find that the in-gap states in \Fref{fig:0h0hsil}(b) are completely different from the flat band in \Fref{fig:0h0hsil}(a). 
In addition, $|\Psi({\alpha})|^{2}$ for the in-gap states at $k_{y}=\pi$ in \Fref{fig:0h0hsil}(c) 
and $k_{y}=0$ in \Fref{fig:0h0hsil}(d) have different orbital components. 
The in-gap states at $k_{y}=\pi$ 
mainly consist of  $p_z$ orbitals localized at the outermost edge sites. 
On the other hand, 
the in-gap states 
at $k_y=0$ mainly consist of  $p_x$ orbitals localized at the outermost edge sites 
and $p_z$ orbitals 
localized at the first neighbor to the outermost edge sites. 
This feature is different from 
the in-gap states for ZGrNRs without hydrogen termination in 
\Fref{fig:0h0hgra}(d). 

\begin{figure}[htbp]
	\includegraphics[width=7.0cm]{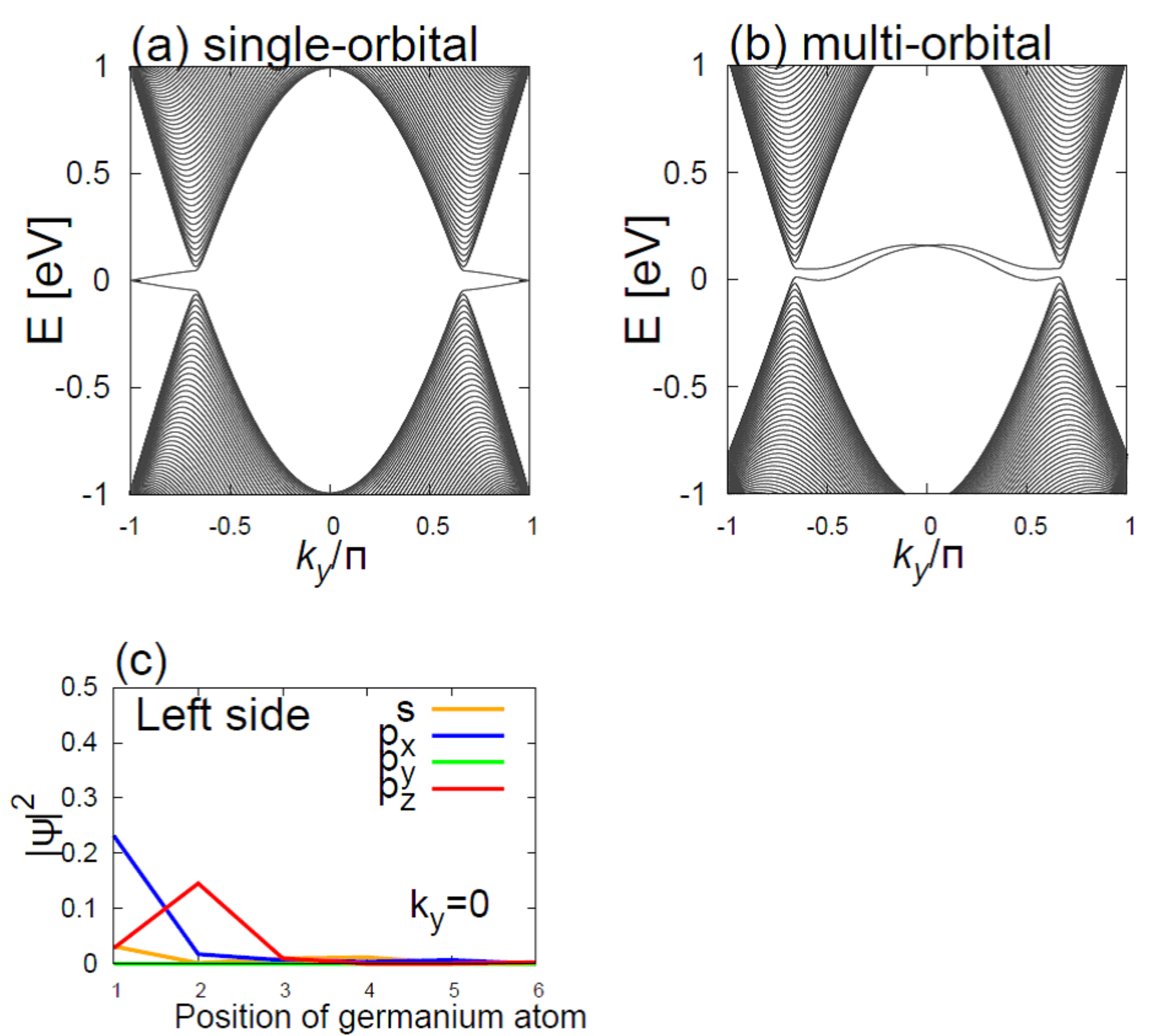}
	\caption{\label{fig:0h0hger} Energy spectra for ZGeNRs ($w=100$) calculated by (a) the single-orbital model and (b) the multi-orbital model (0H/0H). (c) Orbital decomposed probability density 
of edge states  at $k_{y}=0$ in 0H/0H ZGeNRs.}
\end{figure}

Next, we show the energy spectra for ZGeNRs in \Fref{fig:0h0hger}. 
Comparing the results of the single-orbital model (\Fref{fig:0h0hger}(a)) and the multi-orbital model (\Fref{fig:0h0hger}(b)),
the dispersion of the in-gap states is completely different
while the bulk energy spectra are similar, as in the case of ZSiNRs.
To be specific, in the case of the single-orbital model, 
the upper and lower branches of the resulting in-gap states show a linear dispersion 
and cross each other at $k_y=\pm\pi$ (\Fref{fig:0h0hger}(a)) 
where time-reversal invariance is satisfied. 
A bulk energy gap opens at $k_{y}=\pm2\pi/3$ due to the 
spin--orbit coupling. 
This in-gap state can be regarded as  a typical helical 
edge state realized in  a QSH insulator. 
However, the qualitative features of the in-gap state  
in the multi-orbital model are different from that for the single-orbital model. 
As shown in \Fref{fig:0h0hger}(b),   
the in-gap states for $-2\pi/3 \leq k_{y} \leq 2\pi/3$ connect between the conduction and valence bands and  can be regarded as helical edge states. 
Although the present in-gap state is located around the Fermi level $E=0$, 
it is not symmetric around $E=0$, in contrast to the case for 
the single-orbital model.  
Also, the line shapes and positions of the in-gap states for 
ZGeNRs are different from those for ZSiNRs. 
We find from \Fref{fig:0h0hger}(c) that $|\Psi({\alpha})|^{2}$ for the in-gap states at $k_{y}=0$ 
mainly consists of $p_x$ orbitals at the outermost edge sites 
and $p_z$ orbitals at the first neighbor to the outermost edge sites, as shown 
in \Fref{fig:0h0hger}(c). 

\begin{figure}[htbp]
	\includegraphics[width=7cm]{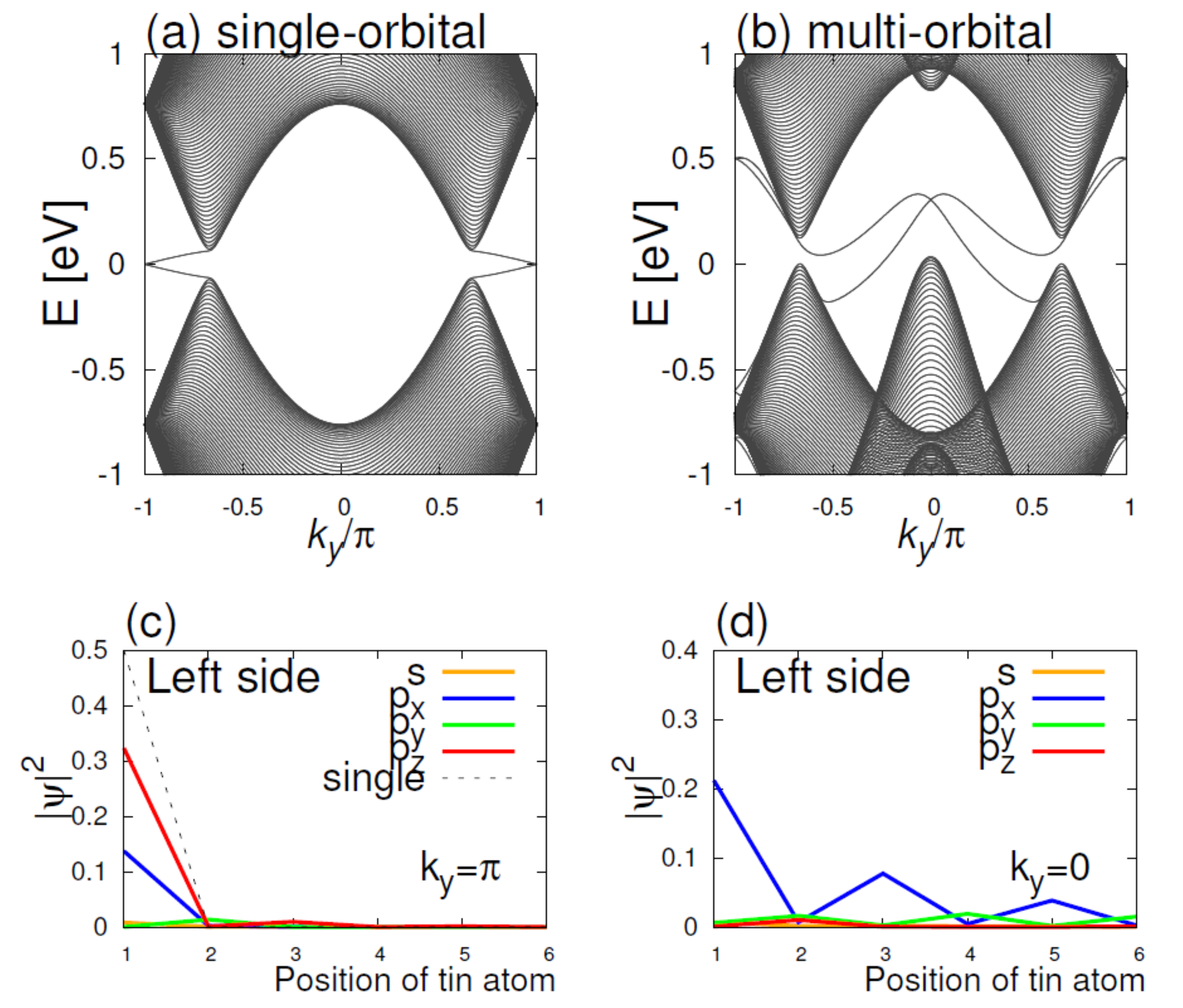}
	\caption{\label{fig:0h0hsta} Energy spectra for ZSnNRs ($w=100$) calculated by (a) the single-orbital model and (b) the multi-orbital model (0H/0H). Orbital decomposed probability density 
of the edge states  at (c) $k_{y}=\pi$, (d) $k_{y}=0$ in 0H/0H ZSnNRs.}
\end{figure}

Finally, we focus on the stanene case.
Comparing the results of the single-orbital model (\Fref{fig:0h0hsta}(a)) and the multi-orbital model (\Fref{fig:0h0hsta}(b)),
we find that the dispersions of the in-gap states are completely different
while the bulk energy spectra are similar except for the extra bands originating from the $\sigma$-orbitals around the $\Gamma$ point ($k_y=0$). 
In the case of the single-orbital model, similar to the ZGeNRs case,
 the in-gap state appears as the edge state of ZSnNRs, as shown in \Fref{fig:0h0hsta}(a). 
The in-gap states with linear dispersion appear around the zone boundary. 
The bulk energy gap opens at $\pm 2\pi/3$ due to the spin--orbit interaction. 
On the other hand, 
the calculated energy dispersion based on the multi-orbital model 
is very different from that based on the single-orbital model, as shown in \Fref{fig:0h0hsta}(b). 
Two kinds of in-gap states with nonlinear dispersion 
appear at $2\pi/3 \leq |k_{y}| \leq \pi$ and $-2\pi/3 \leq k_y \leq 2\pi/3$, and the upper and lower branches cross at $k_y=\pm\pi$ and $k_{y}=0$.  
These in-gap states are expected to be helical edge states 
and the dispersion of these in-gap states 
is distinct from that in the single-orbital model. 
On the other hand, 
we cannot easily compare our obtained energy dispersion (\Fref{fig:0h0hsta}(b)) 
and first-principles calculations \cite{stanene_wo_soc} 
where the spin--orbit coupling is not taken into account. 
We also note that  the bulk energy spectrum with continuum energy levels 
is generated around $E=0$ at $\Gamma$ point, as shown in \Fref{fig:0h0hsta}(b). 
This energy spectrum  is unique to  ZSnNRs. 
It is relevant that the value of $V_{pp\sigma}$ for stanene is 
smaller than that for other materials. 
As shown in \Fref{fig:0h0hsta}(c), $|\Psi(\alpha)|^{2}$ for in-gap states at $k_y=\pi$ 
mainly consists of $p_z$ orbitals at the outermost edge sites 
and $p_y$ orbitals at the first neighbor  to the edge sites. 
On the other hand, $|\Psi(\alpha)|^{2}$ for the in-gap states at $k_{y}=0$ 
mainly consists of $p_x$ orbitals localized  at  the edge sites 
and $p_y$ orbitals at the inner sites adjacent to this edge sites, as shown in 
\Fref{fig:0h0hsta}(d). 
Only in the case of ZSnNRs 
do the $p_{x}$ and $p_{y}$ orbital components of the in-gap state 
penetrate into  the inner sites. 
This is because   
the energy dispersion for the in-gap states at $k_{y}=0$ has a clear slope 
while those for ZSiNRs and ZGeNRs are almost flat. 

In summary, we have calculated energy spectra for ZNRs without hydrogen termination using the multi-orbital model and 
compared them with those  generated by the single-orbital model. 
In the case of ZGrNRs, it is possible to describe a zigzag edge state by the single-orbital model  
since the Hamiltonian of GrNRs in the  multi-orbital model can  
be decomposed into two submatrices of  $\sigma$ and $\pi$ orbitals. 
On the other hand, 
the dispersion of the in-gap states for ZSiNRs, ZGeNRs and ZSnNRs 
with low buckled structures cannot be described by the single-orbital model
even though it can express the bulk energy spectra. 
The multi-orbital Hamiltonians for ZSiNRs, ZGeNRs and ZSnNRs cannot be block diagonalized 
due to $sp^{3}$-like hybridization. 
Therefore, the multi-orbital effects are needed to describe the edge states 
of low buckled material nanoribbons.
Thus, we use the multi-orbital model in the following subsections. 

\subsection{Hydrogen termination effects to ZNRs}
In this subsection, 
we calculate the energy spectra for ZSiNRs, ZGeNRs and ZSnNRs 
with two types of hydrogen termination, the 1H/1H and 2H/2H cases, with a ribbon width $w=100$. 
For ZGrNRs, in an actual experimental situation 
the dangling bond states are terminated by a hydrogen atom and 
the non-zero edge states in \Fref{fig:0h0hgra}(b) disappear. 

\begin{figure}[htbp]%
	\includegraphics[width=8.5cm]{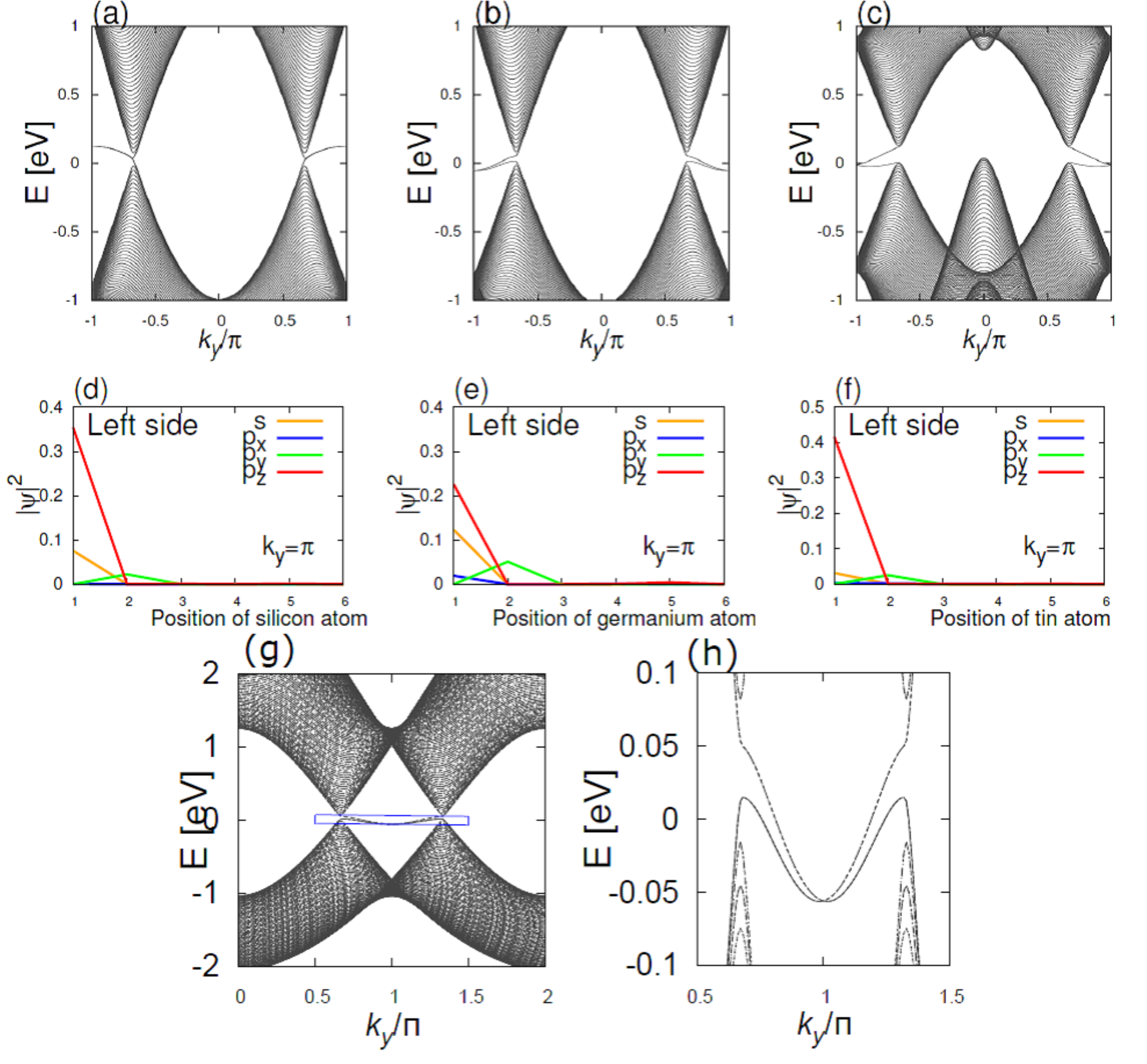}
	\caption{\label{fig:1h1h} Energy spectra for 1H/1H (a) ZSiNRs,  (b) ZGeNRs and (c) ZSnNRs ($w=100$). Orbital decomposed probability density 
of edge states  at $k_{y}=\pi$ in (d) 1H/1H ZSiNRs, (e) ZGeNRs and (f) ZSnNRs. 
(g) Energy dispersion for edge states of 1H/1H ZGeNRs ($w=100$) plotted at $0\leq k_y \leq 2\pi$. (h) Enlarged figure of blue frame in (g).}%
\end{figure}

First, the energy spectra for 1H/1H ZNRs are shown in Figures \ref{fig:1h1h}(a)--(c). 
The in-gap states exist at $2\pi/3 \leq |k_y| \leq \pi$ for all materials and they are helical edge states. 
With an increase in atomic number, the magnitude of the 
bulk energy gap at $k_{y}=\pm 2\pi/3$ increases. 
It becomes the most prominent for ZSnNRs in \Fref{fig:1h1h}(c),  
due to the spin--orbit interaction. 
In addition, the magnitude of the spin splitting of the edge states also increases with the opening of the 
bulk energy gap. 
In the case of 1H/1H ZGeNRs (\Fref{fig:1h1h}(b)), 
a similar energy dispersion of the edge states is obtained by 
first-principles calculations \cite{fujitager}. 
Especially, Figures \ref{fig:1h1h}(g) and (h) are qualitatively the same as those determined by first-principles calculations. 
In addition, the energy dispersion for edge states is non-linear in contrast to the single-orbital model \cite{Ezawa_externalfield}.  
For 1H/1H ZSiNRs and ZSnNRs, first-principles calculations have been performed 
without spin--orbit coupling \cite{silicon_nanoribbons,hydrosil_nm,stanene_wo_soc}. 
Since the spin--orbit interaction for a silicon atom is weak, the obtained energy dispersion for ZSiNRs (\Fref{fig:1h1h}(a)) is close to 
that determined by first-principles calculations \cite{silicon_nanoribbons,hydrosil_nm}. 
On the other hand, since the spin--orbit interaction for a tin atom is strong, 
we cannot easily compare the energy dispersion for 1H/1H ZSnNRs using the multi-orbital model (\Fref{fig:1h1h}(c)) 
and that by first-principles calculations \cite{stanene_wo_soc}. 
$|\Psi(\alpha)|^{2}$ for the edge states at $k_y= \pm \pi$ is shown in Figures \ref{fig:1h1h}(d)--(f). 
$|\Psi(\alpha)|^{2}$ mainly consists 
of $p_{z}$ orbitals localized at the outermost edge in all cases. 
The contributions of the $s$ and $p_{y}$ orbitals are also enhanced near the 
outermost edge sites.  
The major contribution from the $p_z$ orbital 
is a common feature shown in the zigzag edge states in 1H/1H ZGrNRs.  
Therefore, we can qualitatively regard the mono-hydrogen terminated edge states for ZNRs as zigzag-like edge states. 

\begin{figure}[htbp]
	\includegraphics[width=8.5cm]{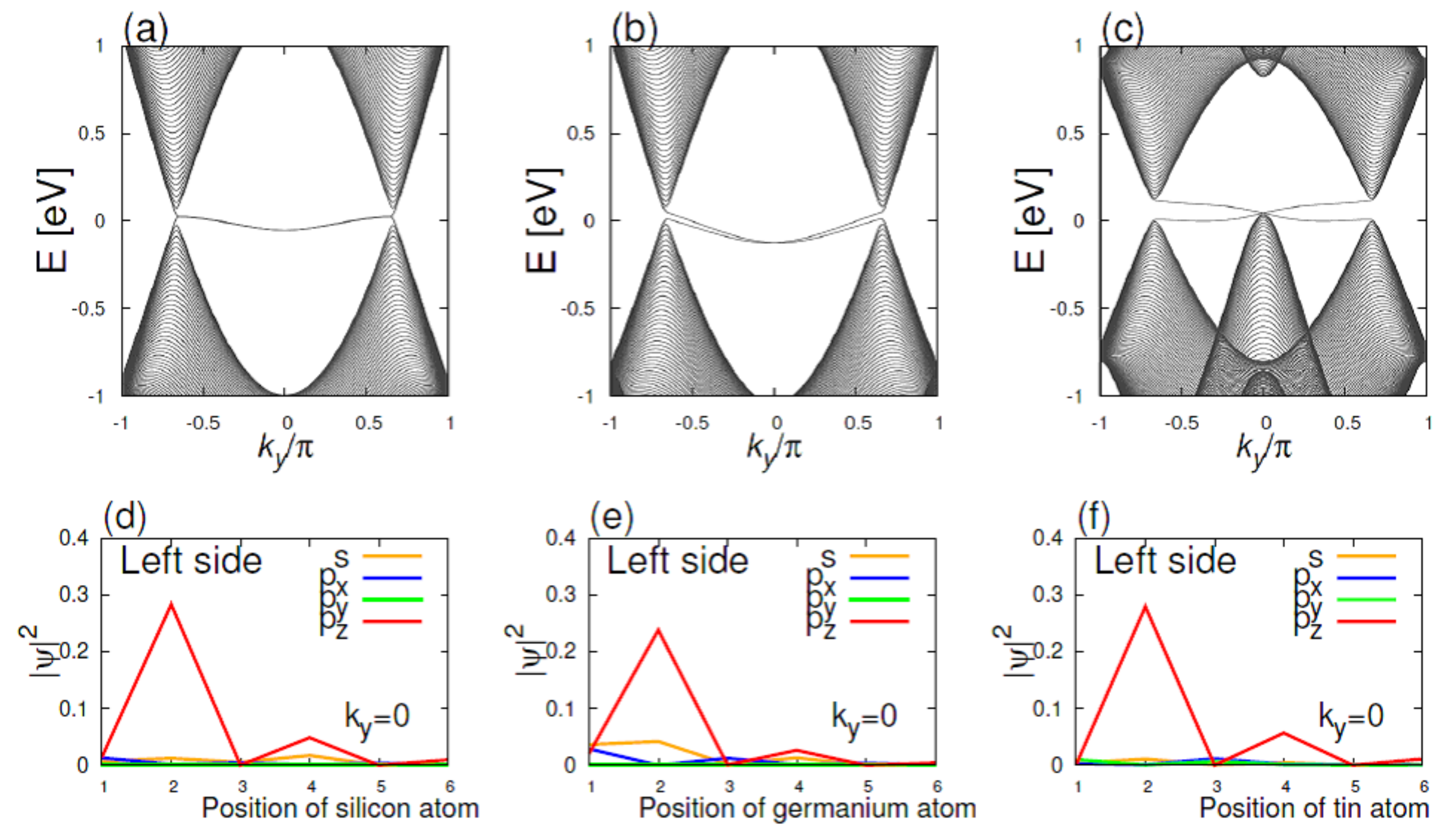}
	\caption{\label{fig:2h2h} Energy spectra for 2H/2H (a) ZSiNRs, (b) ZGeNRs and (c) ZSnNRs ($w=100$). Orbital decomposed probability density 
of edge states at $k_{y}=0$ in 2H/2H (d) ZSiNRs, (e) ZGeNRs and (f) ZSnNRs.}
\end{figure}

Next, the energy spectra for 2H/2H ZNRs are shown in 
Figures \ref{fig:2h2h}(a)--(c). 
The edge states appear at  $-2\pi/3 \leq k_y \leq 2\pi/3$, in contrast to the 
case for 1H/1H  ZNRs. 
With an increase in the spin--orbit coupling, the 
magnitude of the splitting of the edge states becomes prominent, 
as in the case of 1H/1H ZNRs. 
The energy dispersion for the edge states of 2H/2H ZSiNRs is almost 
two-fold degenerate, similar to 1H/1H ZSiNRs as shown in (a). 
The location of the edge states is consistent with 2H/2H ZSiNRs 
in nonmagnetic states by first-principles calculations \cite{hydrosil_nm}.
With increasing strength of the spin--orbit coupling,  
we can see spin-split edge states as helical edge states in 
(b) 2H/2H ZGeNRs and (c) 2H/2H ZSnNRs. 
We note that an edge mode with a crossing point at 
$k_{y}=0$ is generated. \par
$|\Psi(\alpha)|^{2}$ at $k_y=0$ is plotted for (d) ZSiNRs, (e) ZGeNRs and (f) ZSnNRs in \Fref{fig:2h2h}. 
The edge states mainly consist of $p_z$ orbitals localized at the 
first neighbor to the outermost edge sites. 
These results are essentially the same as the Klein 
edge state realized in 2H/2H ZGrNRs. 

To summarize this subsection, 
$p_z$ orbitals make the main contributions to the edge states for ZNRs 
with various types of  hydrogen termination .
The obtained edge states do not have completely flat bands similar to those for ZGrNRs 
due to the orbital hybridization. 
Moreover, the degree of splitting of the helical edge states 
becomes stronger with increasing spin--orbit coupling.  
However,  
the energy dispersion for the edge states is non-linear due to multi-orbital effects. 
Our results show that 
stanene can be regarded as a 
good example of a QSH system, 
since the bulk energy gap is sufficiently large to be observed experimentally.

\section{Summary and Conclusion \label{conclusion}}
In this paper, we have developed a multi-orbital tight-binding model to study the edge states 
of zigzag silicene, germanene and stanene nanoribbons. In this study, the parameters are  
extracted from the DFT results for simple tetragen molecules. We have shown that the energy 
dispersion by the DFT calculations can be successfully reproduced by our multi-orbital 
tight-binding model using these parameters. The obtained dispersion of the edge states is  
non-linear, similar to the DFT results. On the other hand, the dispersion obtained using the single-orbital 
tight-binding model is always linear due to the chiral symmetry. Therefore, we find that the 
non-linearity comes from the multi-orbital effects, and the results cannot be obtained using 
the single-orbital model but can be obtained using the multi-orbital tight-binding model. 
This multi-orbital tight-binding model can be readily applied to complicated and much larger systems. 

\ack
The authors would like to thank Mr. Hiroki Shirakawa for his contributions. This work was supported by a Grant-in Aid for Scientific Research on Innovative Areas "Topological Material Science" (Grant No. JP15H05855 and No. JP15H05853), a Grant-in-Aid for Challenging Exploratory Research (Grant No. JP15K13498), the Core Research for Evolutional Science and Technology (CREST) of the Japan Science and Technology Corporation (JST) and  
JSPS KAKENHI Grants No. JP25107005, JP16K13845,JP26247064.

\section*{Appendix. Calculation of the eigen energies of the multi-orbital Hamiltonian for silane, germane and stannane}
The basis of the Hamiltonian $\mathcal{H}_{mol}$ is given by 
	\begin{equation}
		\Psi^\dagger=(\psi_s,\psi_{p_x},\psi_{p_y},\psi_{p_z},\phi_0,\phi_1,\phi_2,\phi_3)
	\end{equation}
where $\psi_s$, $\psi_{p_x}$, $\psi_{p_y}$ and $\psi_{p_z}$ 
are the bases of the $s$, $p_{x}$, $p_{y}$ and $p_{z}$  orbitals 
of the center atom. 
$\phi_0$,$\phi_1$,$\phi_2$ and $\phi_3$ are 
$1s$ orbitals of the four hydrogen atoms 
bonding the center atom.
The corresponding Hamiltonian $\mathcal{H}_{mol}$ is expressed by 
an $8\times8$ matrix:
	\begin{eqnarray}
		\mathcal{H}_{mol} \nonumber \\
		=\left(
		\begin{array}{cccccccc}
 			\epsilon_{s}  &  0  &  0  &  0  &  V_{ss\sigma}  &  V_{ss\sigma}  &  V_{ss\sigma}  &  V_{ss\sigma}  \\
			0 &  \epsilon_{p}  &  0  &  0  &  V  &  V  &  -V  &  -V  \\
			0 &  0  &  \epsilon_{p}  &  0  &  V  &  -V  &  V  &  -V  \\
			0 &  0  &  0  &  \epsilon_{p}  &  V  &  -V  &  -V  &  V  \\
			V_{ss\sigma}  & V  &  V  &  V  &  \epsilon_{H}  &  0  &  0  &  0  \\
			V_{ss\sigma}  & V  &  -V  &  -V  &  0  &  \epsilon_{H}  &  0  &  0  \\
			V_{ss\sigma}  & -V  &  V  &  -V  &  0  &  0  &  \epsilon_{H}  &  0  \\
			V_{ss\sigma}  & -V  &  -V  &  V  &  0  &  0  &  0  &  \epsilon_{H}  \\
		\end{array}
		\right), \nonumber \\
		{}
	\end{eqnarray}
with $V=\frac{V_{sp\sigma}}{\sqrt{3}}$.  
$\epsilon_{s}$ and $\epsilon_{p}$ are the on-site energies of the outer shell $s$ orbital and the  $p$ orbital of the center atom,
$\epsilon_{H}$ is the on-site energy of a hydrogen atom.
We solve $\mathcal{H}_{mol}\psi=E\psi$ and 
obtain eight eigen values analytically.

After  unitary transformation using the following unitary matrix,  
	\begin{equation}
		U=U_aU_b,
	\end{equation}

	\begin{equation}
		U_a=
		\left(
		\begin{array}{cc}
			A & 0 \\
			0 & A 
		\end{array}
		\right),
	\end{equation}

	\begin{equation}
		A=
		\left(
		\begin{array}{cccc}
			1 & 0 & 0 & 0 \\
			0 & \frac{1}{\sqrt{3}} & \frac{1}{\sqrt{2}} & \frac{1}{\sqrt{6}} \\
			0 & \frac{1}{\sqrt{3}} & -\frac{1}{\sqrt{2}} & \frac{1}{\sqrt{6}} \\
			0 & \frac{1}{\sqrt{3}} & 0 & -\frac{2}{\sqrt{6}} 
		\end{array}
		\right),
	\end{equation}

	\begin{equation}
		U_b=
		\left(
		\begin{array}{cccc}
			I &  & 0 &  \\
			  & I &   &  \\
			  &   & B &  \\
			 &  0 &  & I 
		\end{array}
		\right),
	\end{equation}

	\begin{equation}
		I=
		\left(
		\begin{array}{cc}
			1 & 0 \\
			0 & 1 
		\end{array}
		\right),
	\end{equation}

	\begin{equation}
		B=
		\left(
		\begin{array}{cc}
			\frac{1}{2} & \frac{\sqrt{3}}{2} \\
			\frac{\sqrt{3}}{2} & -\frac{1}{2} 
		\end{array}
		\right),
	\end{equation}
we change the basis as follows: 
	\begin{eqnarray}
		U^\dagger\mathcal{H}_{\mathrm{mol}}U  \nonumber  \\ 
		=\left(
		\begin{array}{cccccccc}
			\epsilon_{s}  &  0  &  0  &  0  &  2V_{ss\sigma}  &  0  &  0  &  0  \\
			0 &  \epsilon_{p}  &  0  &  0  &  0 &  2V  &  0  & 0  \\
			0 &  0  &  \epsilon_{p}  &  0  &  0  &  0  &  2V  &  0  \\
			0 &  0  &  0  &  \epsilon_{p}  &  0  &  0  &  0  &  2V  \\
			2V_{ss\sigma}  & 0  &  0  &  0  &  \epsilon_{H}  &  0  &  0  &  0  \\
			0  & 2V  &  0  &  0  &  0  &  \epsilon_{H}  &  0  &  0  \\
			0  & 0  &  2V  &  0  &  0  &  0  &  \epsilon_{H}  &  0  \\
			0  & 0  &  0  &  2V  &  0  &  0  &  0  &  \epsilon_{H}  
		\end{array}\label{unimat}
		\right). \nonumber \\
		{}
	\end{eqnarray}
We can transform the basis of equation \eref{unimat} as follows:  
	\begin{eqnarray}
		\left(
		\begin{array}{cc}
			\epsilon_s & 2V_{ss\sigma}  \\
			2V_{ss\sigma} & \epsilon_H 
		\end{array}
		\right)
		\oplus
		\left(
		\begin{array}{cc}
			\epsilon_p & 2V  \\
			2V & \epsilon_H 
		\end{array}
		\right)
		\oplus
		\left(
		\begin{array}{cc}
			\epsilon_p & 2V  \\
			2V & \epsilon_H 
		\end{array}
		\right)
		\nonumber \\ 
		\hspace{50mm}\oplus
		\left(
		\begin{array}{cc}
			\epsilon_p & 2V  \\
			2V & \epsilon_H \\
		\end{array}
		\right).\label{mat28}
	\end{eqnarray}
Consequently, we can obtain the energy levels for silane, germane and stannane analytically.	

\section*{Reference}
%\bibliography{edge_thesis_1}

\begin{thebibliography}{10}
\bibitem{first}
The first-principles calculations were performed by the VASP (Vienna ab-initio
  simulation package) code \cite{f1,f2}, which is based on the density
  functional theory (DFT) with the Perdew-Burke-Enzerhof generalized gradient
  approximation \cite{f3}. The core-valence interactions were described by
  projected-augmented-wave potentials \cite{f4}.
\end{thebibliography}
\providecommand{\newblock}{}

\end{document}